\pdfoutput=1
\documentclass[aps,prd,a4paper,preprint,preprintnumbers,titlepage,
nofootinbib,superscriptaddress]{revtex4-2}

\usepackage[a4paper, hdivide={1.91cm,,1.165cm},
vdivide={1.83cm,,3.0cm}]{geometry}

\usepackage{graphicx}
\graphicspath{{figures/}}

\usepackage[hyperfootnotes=false,colorlinks=true,citecolor=blue,urlcolor=blue]{hyperref}

\usepackage{amsfonts}

\usepackage{amsmath}
\usepackage{amssymb}

\usepackage{bbm}

\usepackage{bbold}

\usepackage{adjustbox}
\usepackage{booktabs}

\usepackage{colortbl}

\usepackage{cancel}

\usepackage{color}
\usepackage{xcolor}

\usepackage{graphicx}
\usepackage{xspace}

\usepackage{units}
\usepackage{slashed}

\usepackage{tensor}

\usepackage{gensymb}

\usepackage{cleveref}

\usepackage{tabularx}
\usepackage{multirow}

\usepackage{setspace}
\usepackage{orcidlink}

\usepackage{caption}
\usepackage{subcaption}

\usepackage{enumerate}
\usepackage{enumitem}

\usepackage[version=3]{mhchem}

\usepackage{lipsum}

\newcommand{{\ia} }{{\i}}
\newcommand{{\Ia} }{{\.I}}

\def\s2tw{{\rm sin ^2 \theta_{W}}}

\def\beq{\begin{equation}}
\def\eeq{\end{equation}}
\def\bea{\begin{eqnarray}}
\def\eea{\end{eqnarray}}

\def\bea{\begin{eqnarray}}
\def\eea{\end{eqnarray}}
\def\beeq{\begin{eqnarray}}
\def\eeeq{\end{eqnarray}}

\def\ba{\begin{array}}
\def\ea{\end{array}}

\def\xis0{{\Xi^{*0}}}

\def\g5{\gamma_5}

\allowdisplaybreaks[1]
\newcommand{\Bc}{B_c}
\newcommand{\Bcs}{B_c^*}
\newcommand{\Pone}{B_c(1P_1)}
\newcommand{\Ptwo}{B_c(1P_1^{\prime})}
\newcommand{\GeV}{\mathrm{GeV}}
\newcommand{\MeV}{\mathrm{MeV}}
\newcommand{\keV}{\mathrm{keV}}

\newcommand{\GGc}{\big\langle \frac{\alpha_s}{\pi}G^2\big\rangle}
\newcommand{\thBc}{46.6}
\newcommand{\GOneP}{21.2^{+2.1}_{-1.9}}

\newcommand{\GOneV}{19.0^{+2.6}_{-2.3}}

\newcommand{\GTwoP}{19.5^{+2.4}_{-2.2}}

\newcommand{\GTwoV}{20.1^{+2.8}_{-2.4}}

\newcommand{\ROne}{1.12^{+0.07}_{-0.07}}

\newcommand{\RTwo}{0.97^{+0.10}_{-0.09}}

\newcommand{\fOneRes}{410^{+17}_{-15}}
\newcommand{\fTwoRes}{155^{+7}_{-6}}

\newcommand{\PCthreeMin}{62}
\newcommand{\PCthreeMax}{86}

\newcommand{\GtwoDoubleMax}{2.1}
\newcommand{\NMC}{3000}

\newcommand{\DsixWidthMax}{0.27}
\newcommand{\DsixNmax}{5}

\newcommand{\gPOneMC}{0.308^{+0.006}_{-0.007}}
\newcommand{\gPTwoMC}{0.281^{+0.009}_{-0.009}}
\newcommand{\gaOneMC}{-1.57^{+0.06}_{-0.06}}
\newcommand{\gbOneMC}{-3.38^{+0.13}_{-0.14}}
\newcommand{\gaTwoMC}{-1.84^{+0.06}_{-0.07}}
\newcommand{\gbTwoMC}{8.57^{+0.36}_{-0.34}}

\newcommand{\uzOneP}{0.464}
\newcommand{\uzTwoP}{0.463}
\newcommand{\uzOneV}{0.469}
\newcommand{\uzTwoV}{0.468}

\newcommand{\GtwoShareAll}{0.9}
\newcommand{\DsixShareAll}{0.15}
\newcommand{\varCplMax}{7}
\newcommand{\varCplMin}{3}

\begin{document}

\title{Radiative decays of the axial-vector $B_c(1P)$ mesons as a probe of $^1P_1$--$^3P_1$ mixing}

\author{T.~M.~Aliev\,\orcidlink{0000-0001-8400-7370}}
\email{taliev@metu.edu.tr}
\affiliation{Department of Physics, Middle East Technical University,
Ankara, 06800, Turkey}

\author{S.~Bilmis\,\orcidlink{0000-0002-0830-8873}}
\email{sbilmis@metu.edu.tr}
\affiliation{Department of Physics, Middle East Technical University,
Ankara, 06800, Turkey}
\affiliation{TUBITAK ULAKBIM, Ankara, 06510, Turkey}

\author{M.~Savci\,\orcidlink{0000-0002-6221-4595}}
\email{savci@metu.edu.tr}
\affiliation{Department of Physics, Middle East Technical University,
Ankara, 06800, Turkey}

\date{\today}

\begin{abstract}
We calculate the radiative decays $\Pone\to\Bc\gamma$, $\Pone\to\Bcs\gamma$, $\Ptwo\to\Bc\gamma$, and
$\Ptwo\to\Bcs\gamma$ of the two axial-vector $B_c(1P)$ mesons within QCD sum rules, with the physical states
constructed from the $^1P_1$ and $^3P_1$ configurations. We obtain
\begin{align*}
\Gamma[\Pone\to\Bc\gamma]&=\GOneP~\keV, &
\Gamma[\Pone\to\Bcs\gamma]&=\GOneV~\keV,\\
\Gamma[\Ptwo\to\Bc\gamma]&=\GTwoP~\keV, &
\Gamma[\Ptwo\to\Bcs\gamma]&=\GTwoV~\keV.
\end{align*}
The two basis-current contributions to the $\Bc\gamma$ and $\Bcs\gamma$ transitions interfere constructively for the
lower state and destructively for the higher state. Both states decay into $\Bc\gamma$ and $\Bcs\gamma$ with comparable rates, with the ratios
$R=\Gamma(\Bc\gamma)/\Gamma(\Bcs\gamma)=\ROne$ and $\RTwo$. These ratios are sensitive
probes of the $^1P_1$--$^3P_1$ mixing, and they enter the interpretation of the $B_c(1P)$ structure observed by
the LHCb Collaboration in the $\Bc\gamma$ mass spectrum.
\end{abstract}

\maketitle

\section{Introduction}
\label{sec:intro}

Radiative transitions of heavy mesons complement mass spectroscopy. They are sensitive to the spin structure of the
states and to the charges of the quarks that emit the photon. The $B_c$ mesons, $\bar bc$, are particularly suited
for such studies. Since the two heavy quarks have different flavors, the $B_c$ mesons cannot annihilate into gluons,
and the states below the threshold for strong decays into a $B$ and a $D$ meson cascade
to the ground state by electromagnetic and hadronic transitions. For the $1P$ states, the radiative transitions dominate
the decays~\cite{Gershtein:1994jw,Fulcher:1998ka,Ebert:2002pp,Godfrey:2004ya,Eichten:2019gig}.

The axial-vector ($J^P=1^+$) states are of special interest in this respect. Since charge conjugation is
not a symmetry of the $\bar bc$ system, the physical states are not pure $^1P_1$ or $^3P_1$
configurations~\cite{Godfrey:2004ya}. Instead, spin-dependent interactions mix these two basis states. The mixing
vanishes for equal quark masses. In the limit $m_b\to\infty$, the physical states have definite total
angular momentum of the charm quark, $\boldsymbol j=\boldsymbol L+\boldsymbol s_c$, with $j=\frac12$ and $j=\frac32$, which corresponds to $\theta=35.3^\circ$ (or
$-54.7^\circ$) in the convention of Ref.~\cite{Godfrey:2004ya}. The model predictions summarized in
Ref.~\cite{LHCb:2025prd} lie mostly between $17^\circ$ and
$36^\circ$~\cite{Gershtein:1994jw,Fulcher:1998ka,Ebert:2002pp,Godfrey:2004ya,Eichten:2019gig,Li:2019tbn,Li:2023wgq}, with
a few values outside this range, and a recent potential-model analysis gives
$5^\circ$--$13^\circ$~\cite{MonyA:2026mtz}.
Since both configurations contribute to the radiative amplitudes, constructive or destructive interference may
occur, making the radiative decays of the two states sensitive to the mixing angle.

Since the observation of the ground state by the CDF Collaboration~\cite{CDF:1998ihx}, the $B_c$ spectrum has
been studied at the LHC, where the radially excited $B_c(2S)$ states were observed by the ATLAS, CMS, and LHCb
Collaborations~\cite{ATLAS:2014lga,CMS:2019uhm,LHCb:2019bem}. Recently, the LHCb Collaboration observed a wide
structure in the $\Bc^+\gamma$ mass spectrum, compatible with the lowest orbitally excited $B_c(1P)^+$ mesons and
described by two peaks at $6704.8\pm5.5\pm2.8\pm0.3~\MeV$ and
$6752.4\pm9.5\pm3.1\pm0.3~\MeV$~\cite{LHCb:2025uce,LHCb:2025prd}. Up to six unresolved lines may contribute to this
structure, because each axial-vector state decays into both $\Bc\gamma$ and $\Bcs\gamma$ and the $^3P_0$ and $^3P_2$ states
decay into $\Bcs\gamma$. Since the photon of the subsequent $\Bcs\to\Bc\gamma$ decay is too soft to be detected, the $\Bcs\gamma$
lines appear in the $\Bc\gamma$ spectrum shifted down by about $m_{\Bcs}-m_{\Bc}$~\cite{LHCb:2025prd}. The ATLAS
Collaboration has recently observed the $\Bcs$ meson and measured this mass difference,
$64.5\pm1.4^{+1.0}_{-1.4}~\MeV$~\cite{ATLAS:2026bcs}. The interpretation of the structure, as in the theory-constrained
fits of Ref.~\cite{LHCb:2025prd}, therefore depends on the ratios of the decay rates of the axial-vector states into
$\Bc\gamma$ and $\Bcs\gamma$, which are determined by their mixing.

Radiative transitions of the $B_c$ mesons have been studied in nonrelativistic, relativized, and screened potential
models~\cite{Gershtein:1994jw,Fulcher:1998ka,Godfrey:2004ya,Eichten:2019gig,Li:2019tbn,Li:2022bre,Li:2023wgq,S:2024wld,Bokade:2025lmn,Patel:2026prg},
in the
relativistic quark model~\cite{Ebert:2002pp}, in a Bethe--Salpeter model~\cite{AbdEl-Hady:2005wtn}, and in a
constituent quark model~\cite{Martin-Gonzalez:2022qwd}. Since these approaches use different mixing angles, their predictions for the decay
pattern of the axial-vector states differ considerably.

We have recently determined the mixing angle of the axial-vector $B_c(1P)$ states from two-point QCD sum rules,
$\theta_{B_c(1P)}=(46.6\pm0.4)^\circ$~\cite{Aliev:2026Bcmix}, considerably larger than the values of most quark
models. An important question is whether this large mixing is reflected in the radiative decays. Since both
physical states are then nearly equal mixtures of the two configurations, their decay patterns are expected to
differ markedly from the quark-model ones, and the radiative decays provide a direct test of the mixing.

The present analysis extends our study of the radiative decays of the heavy-strange axial-vector mesons $D_{s1}$
and $B_{s1}$ into pseudoscalar~\cite{Aliev:2026companion} and vector~\cite{Aliev:2026Dsstar} mesons, in which the
physical states were also constructed from two independent interpolating currents. In those decays the photon is
emitted mainly from the strange quark at large distances, and the transitions were described by light-cone sum
rules with photon distribution amplitudes~\cite{Ball:2002ps}. In the $\bar bc$ system both quarks are heavy, and the
photon couples to them perturbatively~\cite{Aliev:1998mq}.

In this work, we calculate the radiative decays $\Pone\to\Bc\gamma$, $\Pone\to\Bcs\gamma$, $\Ptwo\to\Bc\gamma$, and
$\Ptwo\to\Bcs\gamma$ within three-point QCD sum rules, using interpolating currents that mix the $^3P_1$ and
$^1P_1$ configurations at the angle $\theta_{B_c(1P)}$. In the operator product expansion, we take into account the
perturbative contribution and the nonperturbative contributions up to dimension six. We present the widths and the ratios
$R_i=\Gamma(A_i\to\Bc\gamma)/\Gamma(A_i\to\Bcs\gamma)$, which enter the analysis of the LHCb spectrum.

The paper is organized as follows. In Sec.~\ref{sec:framework}, we introduce the interpolating currents and
derive the sum rules. In Sec.~\ref{sec:numerics}, we present the numerical analysis and compare our results
with other approaches. Our conclusions are summarized in
Sec.~\ref{sec:conclusion}. The perturbative spectral densities and the two-point functions are collected in the
Appendices.

\section{Theoretical framework}
\label{sec:framework}

We consider the radiative transitions
\begin{equation}
A_i(p',\eta)\to P(p)\gamma(q,\varepsilon), \qquad
A_i(p',\eta)\to V(p,\tilde\eta)\gamma(q,\varepsilon), \qquad
p'=p+q, \qquad i=1,2,
\label{eq:proc}
\end{equation}
where $A_1\equiv\Pone$ and $A_2\equiv\Ptwo$ are the physical axial-vector mesons, $P=\Bc$, and $V=\Bcs$. Here,
$\eta$, $\tilde\eta$, and $\varepsilon$ denote the polarization vectors of the axial-vector meson, the vector
meson, and the photon, respectively. The meson masses are denoted by $m_{A_i}$, $m_P$, and $m_V$.

The transitions $A_i\to P\gamma$ are described by the vacuum correlation function
\begin{equation}
\Pi^{(i)}_{\mu\nu}(p,q)=i^2\int d^4x\int d^4y\,e^{ip\cdot x+iq\cdot y}
\langle0|T\big\{J^{(i)}_\mu(0)\,J^{\rm em}_\nu(y)\,J_5^\dagger(x)\big\}|0\rangle ,
\label{eq:corr}
\end{equation}
and the transitions $A_i\to V\gamma$ by the function $\Pi^{(i)}_{\mu\nu\alpha}(p,q)$ obtained from
Eq.~\eqref{eq:corr} by replacing $J_5$ with $J^V_\alpha$. Here
\begin{equation}
J^{\rm em}_\nu=e_b\,\bar b\gamma_\nu b+e_c\,\bar c\gamma_\nu c
\label{eq:jem}
\end{equation}
is the electromagnetic current, with $e_b=-\frac13$ and $e_c=\frac23$ the electric charges of the $b$ and $c$ quarks in units of the positron charge, and
the $\Bc$ and $\Bcs$ mesons are interpolated by the currents
\begin{equation}
J_5=\bar b\,i\gamma_5c, \qquad J^V_\alpha=\bar b\gamma_\alpha c .
\label{eq:jPV}
\end{equation}
For the axial-vector channel, we employ the two independent interpolating currents of
Ref.~\cite{Aliev:2026Bcmix},
\begin{equation}
\begin{aligned}
J^A_\mu&=\bar b\gamma_\mu\gamma_5c,\\
J^B_\mu(x)&=-\frac{1}{m_b+m_c}\,\partial^\nu\big[\bar b(x)\,\sigma_{\mu\nu}\gamma_5c(x)\big].
\end{aligned}
\label{eq:basis}
\end{equation}
Here $\sigma_{\mu\nu}=\frac{i}{2}[\gamma_\mu,\gamma_\nu]$, and we use $g_{\mu\nu}={\rm diag}(1,-1,-1,-1)$ and
$\epsilon^{0123}=+1$. The currents $J^A_\mu$ and $J^B_\mu$ correspond to the $^3P_1$ and $^1P_1$ configurations,
respectively. The lower state $A_1$ and the higher state $A_2$ are interpolated by the currents
\begin{equation}
J^{(1)}_\mu=\cos\theta\,J^A_\mu-\sin\theta\,J^B_\mu, \qquad J^{(2)}_\mu=\sin\theta\,J^A_\mu+\cos\theta\,J^B_\mu ,
\label{eq:currents}
\end{equation}
where $\theta$ is the mixing angle.

Following the standard QCD sum-rule procedure, the correlation functions are calculated in two different
representations. On the hadronic side, they are expressed in terms of physical hadronic states, whereas on the QCD
side they are evaluated in terms of quark and gluon degrees of freedom.

The hadronic representation is obtained by inserting complete sets of states with the same quantum numbers as the
interpolating currents and isolating the ground-state contributions. This gives
\begin{equation}
\Pi^{(i)}_{\mu\nu}=\frac{\langle0|J^{(i)}_\mu|A_i(p',\eta)\rangle\langle A_i(p',\eta)|J^{\rm em}_\nu|P(p)\rangle
\langle P(p)|J_5^\dagger|0\rangle}{(m_{A_i}^2-p'^2)(m_P^2-p^2)}+\cdots,
\label{eq:hadronic}
\end{equation}
and the analogous expression for $\Pi^{(i)}_{\mu\nu\alpha}$, where the ellipsis denotes contributions from higher
resonances and continuum states. The relevant matrix elements are
\begin{equation}
\langle0|J^{(i)}_\mu|A_i(p',\eta)\rangle=m_{A_i}f_i\eta_\mu, \qquad
\langle P(p)|J_5^\dagger|0\rangle=\frac{m_P^2f_P}{m_b+m_c}, \qquad
\langle V(p,\tilde\eta)|J^{V\dagger}_\alpha|0\rangle=m_Vf_V\tilde\eta^*_\alpha ,
\label{eq:residues}
\end{equation}
where $f_i$, $f_P$, and $f_V$ denote the decay constants. Lorentz covariance, parity, and electromagnetic gauge
invariance allow the transition matrix elements to be parametrized as
\begin{align}
\langle A_i(p',\eta)|J^{\rm em}_\nu|P(p)\rangle&=g_i\big[(p\cdot q)\eta^*_\nu-(\eta^*\cdot q)p_\nu\big],
\label{eq:gP}\\
\langle A_i(p',\eta)|J^{\rm em}_\nu|V(p,\tilde\eta)\rangle&=g^{(i)}_1\,\epsilon_{\rho\sigma\nu\tau}\eta^{*\rho}\tilde\eta^\sigma q^\tau
+\frac{g^{(i)}_2}{m_{A_i}^2}\,(\eta^*\cdot q)\,\epsilon_{\sigma\nu\rho\tau}\tilde\eta^\sigma p^\rho q^\tau .
\label{eq:gV}
\end{align}
Our aim is to determine the couplings $g_i$, $g^{(i)}_1$, and $g^{(i)}_2$ at $q^2=0$.

After summing over the polarization states and contracting with the photon polarization vector $\varepsilon^\nu$,
we derive the sum rule for $g_i$ from the coefficient of the Lorentz structure
$(p\cdot\varepsilon)q_\mu-\varepsilon_\mu(p\cdot q)$. We do not consider the
structures proportional to $p'_\mu$, since the pseudoscalar meson also contributes to them through
$\langle0|J^A_\mu|P(p')\rangle\propto p'_\mu$. For $A_i\to V\gamma$ we use the coefficients of the structures
\begin{equation}
T_1=\epsilon_{\mu\alpha\rho\sigma}q^\rho\varepsilon^\sigma, \qquad
T_2=q_\mu\,\epsilon_{\alpha\lambda\rho\sigma}p^\lambda q^\rho\varepsilon^\sigma ,
\label{eq:T12}
\end{equation}
which are gauge invariant and free of the double poles of the pseudoscalar state coupled to $J^A_\mu$ and of the
scalar state coupled to $J^V_\alpha$~\cite{Aliev:2026Dsstar}. In the hadronic representation their coefficients are
$-g^{(i)}_1$ and $g^{(i)}_2/m_{A_i}^2$, multiplied by $m_{A_i}f_im_Vf_V/[(m_{A_i}^2-p'^2)(m_V^2-p^2)]$.

We now turn to the calculation of the correlation functions on the QCD side. They are evaluated in the deep
Euclidean region, $p^2,\,p'^2\ll(m_b+m_c)^2$, where the operator product expansion (OPE) is applicable. Since both
quarks are heavy, all propagators are far off shell in this region, and the photon is emitted from the $b$ or the
$c$ quark at short distances, without a long-distance contribution. Applying Wick's theorem to Eq.~\eqref{eq:corr}, we obtain
\begin{equation}
\begin{aligned}
\Pi^{(i)}_{\mu\nu}=\int d^4x\int d^4y\,e^{ip\cdot x+iq\cdot y}
\Big\{&e_c\,{\rm Tr}\big[\Gamma^{(i)}_\mu S_c(-y)\gamma_\nu S_c(y-x)\,i\gamma_5\,S_b(x)\big]\\
&+e_b\,{\rm Tr}\big[\Gamma^{(i)}_\mu S_c(-x)\,i\gamma_5\,S_b(x-y)\gamma_\nu S_b(y)\big]\Big\},
\end{aligned}
\label{eq:wick}
\end{equation}
where $\Gamma^{(i)}_\mu$ denotes the Dirac structure associated with $J^{(i)}_\mu$, $S_Q$ is the propagator of the
heavy quark $Q=b,c$, and the trace is taken over Dirac and color indices. For $\Pi^{(i)}_{\mu\nu\alpha}$, the matrix
$i\gamma_5$ is replaced by $\gamma_\alpha$. The two terms describe the photon emission from the $c$ and the $b$
quark, respectively.

For the propagators we use the heavy-quark propagator in the background gluon field in the Fock--Schwinger
gauge~\cite{Aliev:2026Bcmix,Novikov:1983gd,Reinders:1984sr}, which contains terms with up to three gluon field strengths and with covariant derivatives of the field strength.
For the three-point function the gauge condition, $(z-x_0)^\mu A^a_\mu(z)=0$, refers to a fixed point $x_0$, and the
gluon field at a point $z$ of a quark line is
\begin{equation}
A^a_\mu(z)=\sum_{r=0}^{\infty}\frac{1}{r!\,(r+2)}\,(z-x_0)^\nu(z-x_0)^{\alpha_1}\cdots(z-x_0)^{\alpha_r}\,
D_{\alpha_1}\cdots D_{\alpha_r}G^a_{\nu\mu}(x_0),
\label{eq:FS}
\end{equation}
with $D_\mu=\partial_\mu-ig_st^aA^a_\mu$. The dimension-four contribution contains two insertions of the $r=0$ term
of Eq.~\eqref{eq:FS}, and the dimension-six contributions contain three insertions of the $r=0$ term, two insertions of the $r=1$ term, or one insertion of the
$r=0$ and one of the $r=2$ term. All distributions of the insertions over the three quark lines are included. The
vacuum averages are
\begin{equation}
\begin{aligned}
\langle0|G^a_{\alpha\beta}G^b_{\mu\nu}|0\rangle&=\frac{\delta^{ab}}{96}\langle G^2\rangle\,
\mathcal G_{\alpha\beta,\mu\nu},\\
\langle0|g_s^3G^a_{\mu\nu}G^b_{\rho\sigma}G^c_{\alpha\beta}|0\rangle&=\frac{f^{abc}}{576}\,\langle g_s^3fG^3\rangle\,
T_{\mu\nu,\rho\sigma,\alpha\beta},\\
\langle0|g_s^2G^a_{\mu\nu}D_\rho D_\sigma G^b_{\alpha\beta}|0\rangle&=\frac{\delta^{ab}}{8}
\Big\{\frac{\langle g_s^3fG^3\rangle}{48}\big[2g_{\rho\sigma}\mathcal G_{\mu\nu,\alpha\beta}
+T_{\mu\nu,\rho\sigma,\alpha\beta}\big]
+\frac{\langle g_s^4j^2\rangle}{72}\big[-2g_{\rho\sigma}\mathcal G_{\mu\nu,\alpha\beta}
+T'_{\mu\nu,\rho\sigma,\alpha\beta}\big]\Big\},
\end{aligned}
\label{eq:vac}
\end{equation}
and $\langle0|g_s^2D_\rho G^a_{\mu\nu}D_\sigma G^b_{\alpha\beta}|0\rangle=
-\langle0|g_s^2G^a_{\mu\nu}D_\rho D_\sigma G^b_{\alpha\beta}|0\rangle$, where
$\mathcal G_{\mu\nu,\alpha\beta}=g_{\mu\alpha}g_{\nu\beta}-g_{\mu\beta}g_{\nu\alpha}$,
\begin{equation}
T_{\mu\nu,\rho\sigma,\alpha\beta}=\mathcal G_{\mu\nu,\rho\alpha}g_{\sigma\beta}
-\mathcal G_{\mu\nu,\rho\beta}g_{\sigma\alpha}-\mathcal G_{\mu\nu,\sigma\alpha}g_{\rho\beta}
+\mathcal G_{\mu\nu,\sigma\beta}g_{\rho\alpha},
\label{eq:Tvac}
\end{equation}
and $T'$ differs from $T$ by the signs of the last two terms. Here
$\langle G^2\rangle=\langle G^a_{\mu\nu}G^{a\,\mu\nu}\rangle$, $\langle g_s^3fG^3\rangle=\langle g_s^3f^{abc}G^a_{\mu\nu}G^{b\,\nu}{}_{\lambda}G^{c\,\lambda\mu}\rangle$
and $\langle g_s^4j^2\rangle=g_s^4\langle j^a_\mu j^{a\mu}\rangle$ with $j^a_\mu=\sum_{q=u,d,s}\bar q\gamma_\mu t^aq$.
The last relation of Eq.~\eqref{eq:vac} follows from the Bianchi identity and the gluon equation of motion,
$D^\mu G^a_{\mu\nu}=-g_sj^a_\nu$~\cite{Nikolaev:1982rq,CostaiReina:2026bc}. Through it, the light quarks enter the
correlation functions only via the four-quark operator $\langle g_s^4j^2\rangle$.

The OPE representation of each invariant amplitude can therefore be written as
\begin{equation}
\Pi^{\rm OPE}_X=\Pi^{\rm pert}_X+\Pi^{\langle G^2\rangle}_X+\Pi^{\langle G^3\rangle}_X+\Pi^{\langle j^2\rangle}_X,
\qquad X=A,B,
\label{eq:ope}
\end{equation}
where $X$ labels the axial-vector basis current. The light-quark condensate does not contribute,
and the heavy-quark condensate is small, since $\langle\bar QQ\rangle=-\GGc/(12m_Q)+\cdots$~\cite{Shifman:1978bx}.
Its contribution is contained in the gluon-condensate terms. The interpolating currents of the physical states are
linear combinations of the basis currents, so that the corresponding invariant amplitudes are
\begin{equation}
\widehat\Pi^{(1)}=\cos\theta\,\widehat\Pi_A-\sin\theta\,\widehat\Pi_B, \qquad
\widehat\Pi^{(2)}=\sin\theta\,\widehat\Pi_A+\cos\theta\,\widehat\Pi_B,
\label{eq:rot}
\end{equation}
for each of the Lorentz structures defined above, where the hats denote the double Borel transforms.

Having determined the coefficients of the selected Lorentz structures on both the hadronic and OPE sides, we equate
the two representations and perform double Borel transformations with respect to $p'^2$ and $p^2$, with parameters
$M_1^2$ and $M_2^2$, which suppress the contributions of higher states. Before imposing a relation between the two
Borel parameters, the matching conditions take the form
\begin{equation}
\begin{aligned}
g_i\,m_{A_i}f_i\,\frac{m_P^2f_P}{m_b+m_c}\,e^{-m_{A_i}^2/M_1^2-m_P^2/M_2^2}&=\widehat\Pi^{(i)}(M_1^2,M_2^2,s_{\rm eff}),\\
-g^{(i)}_1\,m_{A_i}f_i\,m_Vf_V\,e^{-m_{A_i}^2/M_1^2-m_V^2/M_2^2}&=\widehat\Pi^{(i)}_{T_1}(M_1^2,M_2^2,s_{\rm eff}),\\
\frac{g^{(i)}_2}{m_{A_i}^2}\,m_{A_i}f_i\,m_Vf_V\,e^{-m_{A_i}^2/M_1^2-m_V^2/M_2^2}&=\widehat\Pi^{(i)}_{T_2}(M_1^2,M_2^2,s_{\rm eff}),
\end{aligned}
\label{eq:match}
\end{equation}
where $\widehat\Pi^{(i)}$, $\widehat\Pi^{(i)}_{T_1}$, and $\widehat\Pi^{(i)}_{T_2}$ are the Borel-transformed coefficients of the structures
$(p\cdot\varepsilon)q_\mu-\varepsilon_\mu(p\cdot q)$, $T_1$, and $T_2$, respectively. It is convenient to introduce the
effective Borel parameter and momentum fractions
\begin{equation}
M^2=\frac{M_1^2M_2^2}{M_1^2+M_2^2}, \qquad u_0=\frac{M_2^2}{M_1^2+M_2^2}, \qquad \bar u_0=1-u_0 .
\label{eq:Meff}
\end{equation}
As in Refs.~\cite{Aliev:2026companion,Aliev:2026Dsstar}, we fix the ratio of the two Borel parameters by the ratio
of the squared masses of the states at the two vertices,
\begin{equation}
\frac{M_1^2}{M_2^2}=\frac{m_{\rm in}^2}{m_{\rm out}^2}, \qquad
u_0=\frac{m_{\rm out}^2}{m_{\rm in}^2+m_{\rm out}^2}, \qquad
M_1^2=\frac{M^2}{u_0}, \qquad M_2^2=\frac{M^2}{\bar u_0},
\label{eq:u0}
\end{equation}
where $m_{\rm in}$ refers to the state at the vertex with momentum $p'$. This gives $u_0=\uzOneP$ and $\uzTwoP$
for $A_{1,2}\to\Bc\gamma$ and $u_0=\uzOneV$ and $\uzTwoV$ for $A_{1,2}\to\Bcs\gamma$. Since the masses of the initial
and final states are close, $u_0$ differs little from $\frac12$.

In terms of the dispersion variables $s_1$ and $s_2$ of the two channels, the Borel exponent is
$s_1/M_1^2+s_2/M_2^2=(u_0s_1+\bar u_0s_2)/M^2$. At $q^2=0$ the double spectral density of the triangle diagram is
concentrated at $s_1=s_2$, and the perturbative amplitudes take the single-dispersion form
\begin{equation}
\widehat\Pi^{\rm pert}_X(M^2,s_{\rm eff})=\int_{(m_b+m_c)^2}^{s_{\rm eff}}ds\,e^{-s/M^2}\rho_X(s), \qquad X=A,B,
\label{eq:pert}
\end{equation}
for every $u_0$, with spectral densities that depend on $u_0$. Here, $s_{\rm eff}$ is the effective threshold of
a triangular duality region, $u_0s_1+\bar u_0s_2\le s_{\rm eff}$, whose intercepts are in the ratio
$M_1^2:M_2^2$~\cite{Pullin:2021ebn}, as in Refs.~\cite{Aliev:2026companion,Aliev:2026Dsstar}.

For photon emission from the $b$ and the $c$ quark, $\rho_X(s)=e_b\,\rho_{X,b}(s)+e_c\,\rho_{X,c}(s)$. For
$A_i\to\Bc\gamma$ we obtain
\begin{equation}
\rho_A(s)=e_b\,F(s;m_b,m_c)-e_c\,F(s;m_c,m_b), \qquad
\rho_B(s)=e_b\,G(s;m_b,m_c)+e_c\,G(s;m_c,m_b),
\label{eq:rhoP}
\end{equation}
with
\begin{equation}
\begin{aligned}
F(s;m_1,m_2)&=\frac{3}{4\pi^2}\Big[m_1L(s;m_1,m_2)-\big((1+\bar u_0)m_1+u_0m_2\big)\frac{\sqrt\lambda}{s}
+\bar u_0(m_1-m_2)^2(m_1+m_2)\frac{\sqrt\lambda}{s^2}\Big],\\
G(s;m_1,m_2)&=\frac{3}{4\pi^2(m_1+m_2)}\Big[m_1(m_2-2\bar u_0m_1)L(s;m_1,m_2)
+\big(m_1^2-m_2^2+(3-4u_0)s\big)\frac{\sqrt\lambda}{2s}\Big],
\end{aligned}
\label{eq:FG}
\end{equation}
where $\lambda\equiv\lambda(s,m_b^2,m_c^2)$, with the Källén function $\lambda(a,b,c)=(a-b-c)^2-4bc$, and
\begin{equation}
L(s;m_1,m_2)=\ln\frac{s+m_1^2-m_2^2+\sqrt\lambda}{s+m_1^2-m_2^2-\sqrt\lambda}.
\label{eq:L}
\end{equation}
The spectral densities for $A_i\to\Bcs\gamma$ are given in
Appendix~\ref{app:pert}.

For the gluon-condensate contributions, after the double Borel transformation, we obtain
\begin{equation}
\widehat\Pi^{\langle G^2\rangle}_X(M^2)=\GGc\sum_{j=b,c}e_j\int_0^1dx\,e^{-S(x)/M^2}
\sum_{n=0}^{3}\frac{C^{(j)}_{X,n}(x)}{(M^2)^n}, \qquad S(x)=\frac{m_c^2}{1-x}+\frac{m_b^2}{x},
\label{eq:G2}
\end{equation}
where the coefficient functions $C^{(j)}_{X,n}(x)$ are rational functions of $x$, $u_0$, and the quark masses. The
dimension-six contributions have the same form, with $\GGc\,C^{(j)}_{X,n}$ replaced by
$\langle g_s^3fG^3\rangle\,D^{(j)}_{X,n}+\langle g_s^4j^2\rangle\,E^{(j)}_{X,n}$ and $n\le\DsixNmax$. The explicit
expressions of these coefficient functions are lengthy, and we do not present them here. As in
Refs.~\cite{Aliev:2026companion,Aliev:2026Dsstar}, the continuum subtraction in the three-point functions is applied
to the perturbative term only.

The sum rules \eqref{eq:match} depend on the residues $f_i$ of the physical states. These are determined from
the two-point correlation functions
\begin{equation}
\Pi^{XY}_{\mu\nu}(p)=i\int d^4x\,e^{ip\cdot x}\langle0|T\big\{J^X_\mu(x)J^{Y\dagger}_\nu(0)\big\}|0\rangle,
\qquad X,Y=A,B,
\label{eq:2pt}
\end{equation}
whose invariant amplitudes give
\begin{equation}
f_i^2m_{A_i}^2e^{-m_{A_i}^2/M^2}=\widehat\Pi_{ii}(M^2,s_0),
\label{eq:fA}
\end{equation}
where $\widehat\Pi_{ii}$ is the Borel-transformed two-point function of the current $J^{(i)}_\mu$, obtained from
$\widehat\Pi_{AA}$, $\widehat\Pi_{AB}$, and $\widehat\Pi_{BB}$ (Appendix~\ref{app:2pt}) by the rotation of
Eq.~\eqref{eq:currents}.

Substituting the residues into Eq.~\eqref{eq:match}, we obtain the sum
rules for the couplings. Since they have the same form for both states, we omit the state index $i$ of the masses,
couplings, and amplitudes, e.g., $m_A\equiv m_{A_i}$. Then
\begin{equation}
\begin{aligned}
g&=\frac{m_b+m_c}{m_Af_i\,m_P^2f_P}\,e^{m_A^2/M_1^2+m_P^2/M_2^2}\,\widehat\Pi(M^2,s_{\rm eff}),\\
g_1&=-\frac{e^{m_A^2/M_1^2+m_V^2/M_2^2}}{m_Af_i\,m_Vf_V}\,\widehat\Pi_{T_1}(M^2,s_{\rm eff}), \qquad
g_2=\frac{m_A\,e^{m_A^2/M_1^2+m_V^2/M_2^2}}{f_i\,m_Vf_V}\,\widehat\Pi_{T_2}(M^2,s_{\rm eff}),
\end{aligned}
\label{eq:gfixed}
\end{equation}
where $M_1^2$ and $M_2^2$ are given by Eq.~\eqref{eq:u0} for each transition. The couplings determine the radiative
decay widths,
\begin{align}
\Gamma(A\to P\gamma)&=\frac{\alpha_{\rm em}}{3}\,k_P^3\,g^2,
\label{eq:widthP}\\
\Gamma(A\to V\gamma)&=\frac{\alpha_{\rm em}}{3}\,k_V^3\Big[\frac{g_1^2}{m_V^2}+\frac{(g_1m_A+g_2k_V)^2}{m_A^4}\Big],
\label{eq:widthV}
\end{align}
where $k_{P,V}=(m_A^2-m_{P,V}^2)/(2m_A)$ is the photon energy and $\alpha_{\rm em}$ is the fine-structure constant.
The two terms of Eq.~\eqref{eq:widthV} are the contributions of the longitudinally and transversely polarized $\Bcs$.

\section{Numerical analysis}
\label{sec:numerics}

The input parameters are collected in Table~\ref{tab:inputs}. Since the masses of the axial-vector states have not been measured, we take the means of the
predictions summarized in Ref.~\cite{LHCb:2025prd}, with an uncertainty of $15~\MeV$ that accounts for their spread.
These masses are compatible with the higher LHCb peak and with the lattice results of Ref.~\cite{Mathur:2018epb}.

\begin{table}[!htbp]
\caption{The values of the input parameters used in the numerical analysis. The quark masses are the
$\overline{\rm MS}$ masses $m_b(m_b)$ and $m_c(m_c)$, which agree with the values $4.186(6)$ and $1.2729(45)~\GeV$ of
Ref.~\cite{ParticleDataGroup:2026} within the uncertainties used here. The four-quark condensate is evaluated with
$\langle\bar qq\rangle(2~\GeV)=-(0.278~\GeV)^3$ for $q=u,d$~\cite{CostaiReina:2026bc,FlavourLatticeAveragingGroupFLAG:2021npn},
$\langle\bar ss\rangle/\langle\bar qq\rangle=0.8$~\cite{CostaiReina:2026bc,Ioffe:2002ee}, and $g_s$ at $\mu=3~\GeV$. The $\Bcs$
mass is $m_{\Bc}$ plus the mass difference measured in Ref.~\cite{ATLAS:2026bcs}.}
\label{tab:inputs}
\begin{ruledtabular}
\begin{tabular}{ll}
Parameter & Value \\
\hline
$m_b$, $m_c$ & $4.18\pm0.03$, $1.27\pm0.02~\GeV$~\cite{ParticleDataGroup:2026} \\
$\langle\alpha_sG^2\rangle$ & $(6.49\pm0.35)\times10^{-2}~\GeV^4$~\cite{Narison:2021gqs} \\
$\langle g_s^3fG^3\rangle$ & $(8.2\pm1.0)~\GeV^2\,\langle\alpha_sG^2\rangle$~\cite{Narison:2021gqs} \\
$\langle g_s^4j^2\rangle=-\frac49r_{\rm vac}\,g_s^4\sum_{q=u,d,s}\langle\bar qq\rangle^2$ &
$-6.5\times10^{-3}\,r_{\rm vac}~\GeV^6$, $r_{\rm vac}=1.0\pm0.5$~\cite{CostaiReina:2026bc,Ioffe:2002ee} \\
$\theta$ & $(46.6\pm0.4)^\circ$~\cite{Aliev:2026Bcmix} \\
$m_{\Bc}$, $m_{\Bcs}$ & $6274.47\pm0.32~\MeV$~\cite{ParticleDataGroup:2026}, $6339.0\pm1.7~\MeV$~\cite{ATLAS:2026bcs} \\
$m_{\Pone}$, $m_{\Ptwo}$ & $6742\pm15$, $6756\pm15~\MeV$ (mean of Ref.~\cite{LHCb:2025prd}) \\
$f_{\Bc}$, $f_{\Bcs}/f_{\Bc}$ & $427(6)~\MeV$~\cite{McNeile:2012qf}, $0.988(27)$~\cite{Colquhoun:2015oha} \\
\end{tabular}
\end{ruledtabular}
\end{table}

For the transition sum rules we use the working region $M^2=10$--$14~\GeV^2$ and
$s_{\rm eff}=54\pm1~\GeV^2$~\cite{Aliev:2026Bcmix} for both axial-vector channels. In
this region, at the central thresholds, the pole contribution of the three-point sum rules is
$\PCthreeMin$--$\PCthreeMax\%$, the gluon condensate contributes at most $\GtwoShareAll\%$ to the couplings, and the
dimension-six terms at most $\DsixShareAll\%$. For the residues of the axial-vector states,
Eq.~\eqref{eq:fA} gives $f_1=\fOneRes~\MeV$ and $f_2=\fTwoRes~\MeV$.

\begin{figure}[!htbp]
\centering
\includegraphics[width=\linewidth]{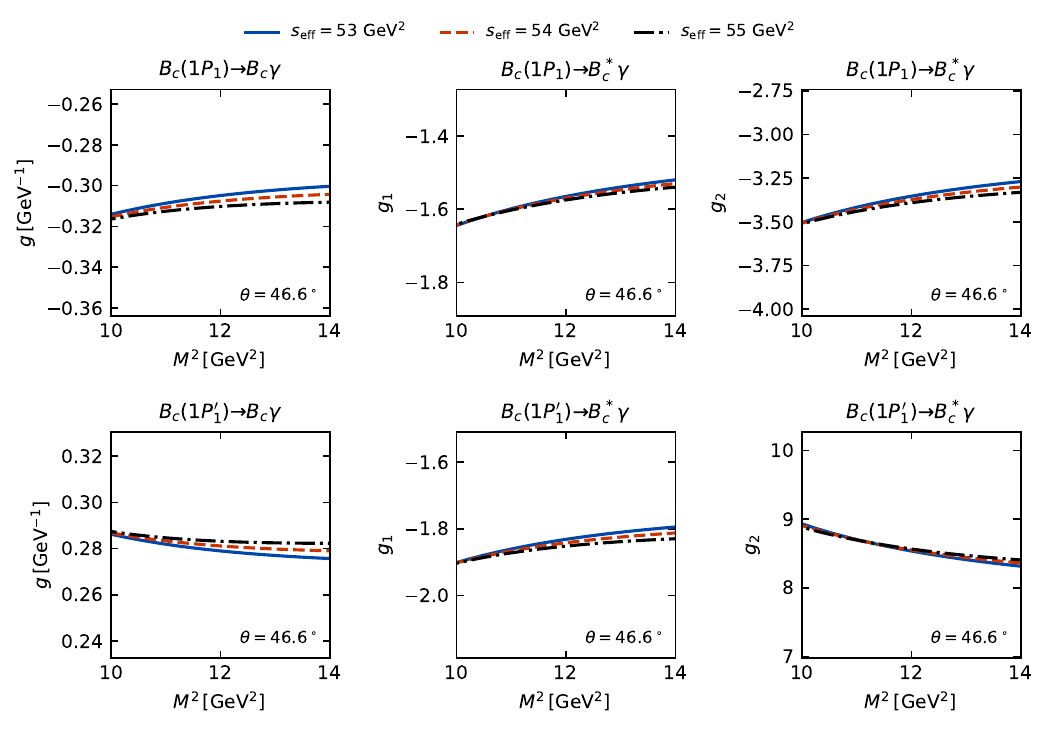}
\caption{Dependence of the couplings $g$ of $A_i\to\Bc\gamma$ (left) and $g_1$ (middle) and $g_2$ (right) of
$A_i\to\Bcs\gamma$ for $\Pone$ (upper panels) and $\Ptwo$ (lower panels) on the Borel parameter $M^2$ at three
values of the effective threshold, $s_{\rm eff}$, with the threshold of the residue sum rules shifted by the same
amount. The Borel prescription of Eq.~\eqref{eq:u0} and the mixing angle $\theta=\thBc^\circ$ are used.}
\label{fig:borel}
\end{figure}

Figure~\ref{fig:borel} shows the dependence of the couplings on $M^2$ at three values of the thresholds. Over the
working region they change by $\varCplMin$--$\varCplMax\%$.

The contributions of $J^A$ and $J^B$ to $g$ and $g_1$ interfere constructively for the lower
state and destructively for the higher state, as for the heavy-strange
mesons~\cite{Aliev:2026companion,Aliev:2026Dsstar}, and the tensor current dominates the amplitude of the higher
state. The axial-vector current does not contribute to the perturbative
part of $g_2$, which is therefore generated by the tensor current and has opposite signs for the two states.

\begin{figure}[!htbp]
\centering
\includegraphics[width=\linewidth]{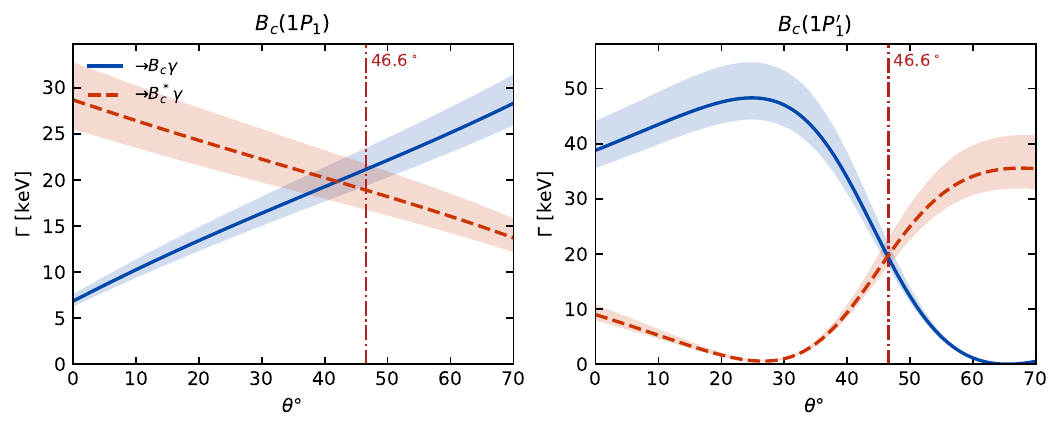}
\caption{Dependence of the radiative widths of $\Pone$ (left) and $\Ptwo$ (right) on the mixing angle, with the
residues $f_i$ recomputed at each angle. The lines show the central values and the shaded bands the central $68\%$
intervals of the Monte Carlo distribution at fixed angle. The dash-dotted line indicates the adopted angle.}
\label{fig:theta}
\end{figure}

\begin{figure}[!htbp]
\centering
\includegraphics[width=0.5\linewidth]{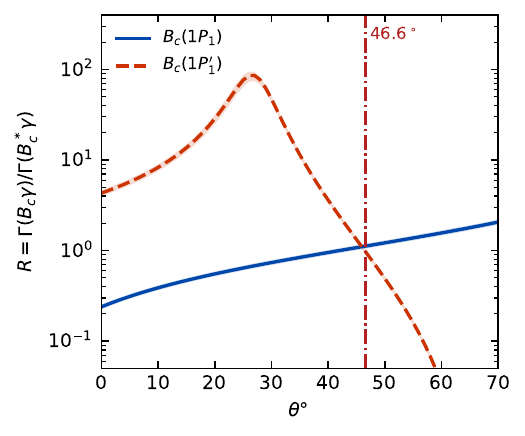}
\caption{Ratios $R_i=\Gamma(A_i\to\Bc\gamma)/\Gamma(A_i\to\Bcs\gamma)$ as functions of the mixing angle, on a
logarithmic scale. Lines and bands as in
Fig.~\ref{fig:theta}.}
\label{fig:ratio}
\end{figure}

The dependence of the widths and of their ratios on the mixing angle is shown in Figs.~\ref{fig:theta} and
\ref{fig:ratio}. For $0^\circ\lesssim\theta\lesssim36^\circ$, which covers most quark models,
our sum rules give the familiar pattern, with the higher state decaying mainly into $\Bc\gamma$ and the lower one
mainly into $\Bcs\gamma$, and $\Gamma(\Ptwo\to\Bcs\gamma)$ nearly vanishes at $\theta\simeq27^\circ$. In the model of
Ref.~\cite{Li:2022bre} the lower state is the $^1P_1$-dominated one, which corresponds to $\theta=65.7^\circ$. At this
angle our sum rules reproduce the reversed pattern of that model, with the lower state decaying mainly into $\Bc\gamma$
and the higher one almost entirely into $\Bcs\gamma$. At $\theta=\thBc^\circ$ the two states are nearly equal mixtures
of $J^A$ and $J^B$, and both decay into $\Bc\gamma$ and $\Bcs\gamma$ with comparable rates, so that each of them
contributes to both lines of the LHCb spectrum. A measurement of the relative yields of the $\Bc\gamma$ and $\Bcs\gamma$
lines of each axial-vector state would therefore test the mixing angle directly.

\begin{figure}[!htbp]
\centering
\includegraphics[width=\linewidth]{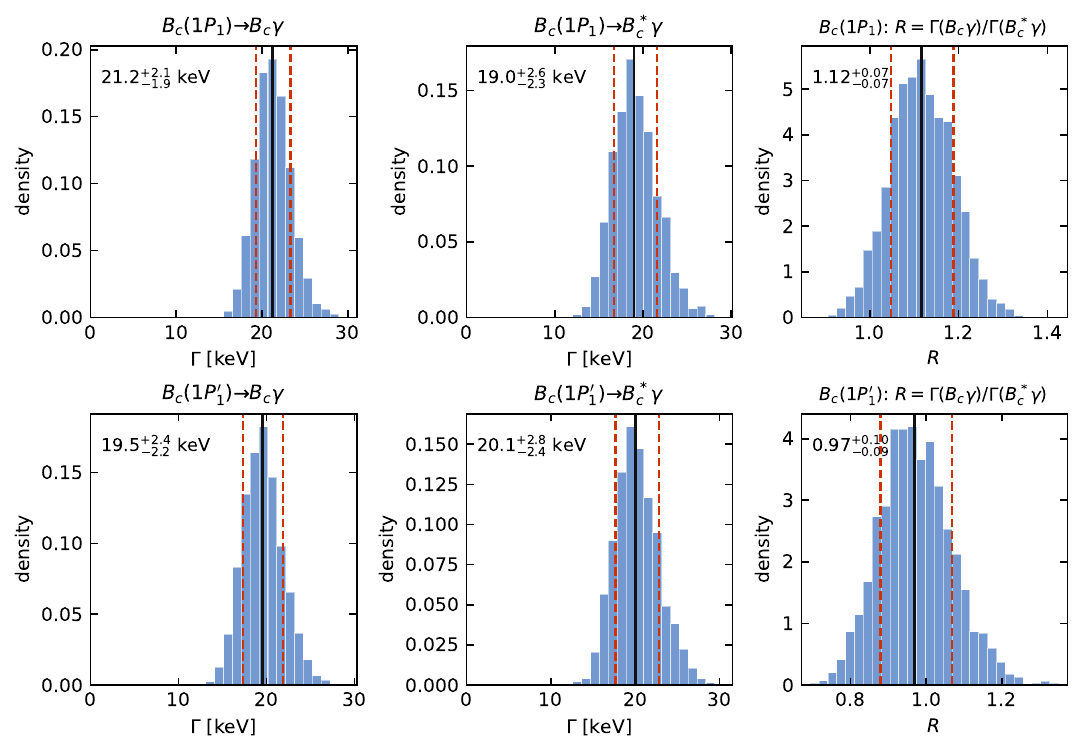}
\caption{Monte Carlo distributions of the radiative widths (left and middle) and of the ratios $R_i$ (right). Solid
vertical lines indicate the medians, and dashed lines the 16th--84th percentiles.}
\label{fig:mc}
\end{figure}

The uncertainties are obtained in a Monte Carlo analysis with $\NMC$ samples, in which the inputs of
Table~\ref{tab:inputs} are varied within their uncertainties, the mass of the lower axial-vector state by $\pm15~\MeV$
and the splitting of the two states by $14\pm10~\MeV$, so that their ordering is preserved, and, simultaneously, the
Borel parameter and the thresholds uniformly within the working region. We quote the medians and the central $68\%$ intervals. The resulting couplings, widths, and ratios are given in Table~\ref{tab:results}, and the distributions in Fig.~\ref{fig:mc}.

\begin{table*}[!htbp]
\caption{Couplings, radiative widths, and their ratio
$R=\Gamma(A_i\to\Bc\gamma)/\Gamma(A_i\to\Bcs\gamma)$. Monte Carlo
medians with central $68\%$ intervals, with $g$ in GeV$^{-1}$ and $g_{1,2}$ dimensionless.}
\label{tab:results}
\begin{ruledtabular}
{\small\setlength{\tabcolsep}{3pt}
\begin{tabular}{lcccccc}
State & $|g|$ & $\Gamma(\Bc\gamma)$ (keV) & $g_1$ & $g_2$ & $\Gamma(\Bcs\gamma)$ (keV) & $R$ \\
\hline
$\Pone$ & $\gPOneMC$ & $\GOneP$ & $\gaOneMC$ & $\gbOneMC$ & $\GOneV$ & $\ROne$ \\
$\Ptwo$ & $\gPTwoMC$ & $\GTwoP$ & $\gaTwoMC$ & $\gbTwoMC$ & $\GTwoV$ & $\RTwo$ \\
\end{tabular}}
\end{ruledtabular}
\end{table*}

At the central values of the parameters, doubling the gluon condensate changes the widths by at most
$\GtwoDoubleMax\%$, and omitting the dimension-six terms in the two- and three-point sum rules changes them by at most
$\DsixWidthMax\%$.

Table~\ref{tab:comparison} compares our results with other predictions. Since our sum rules reproduce the
qualitative patterns of the quark models at their mixing angles, the difference in the decay patterns reflects
the larger mixing angle used here. The model of
Ref.~\cite{S:2024wld}, with $\theta\simeq43^\circ$, finds, as we do, comparable rates into $\Bc\gamma$ and $\Bcs\gamma$
for both states. The sums of
the two widths of each state are smaller than the quark-model values.

\begin{table*}[!htbp]
\caption{Comparison with other predictions, with widths in keV. In all rows $1P_1$ is the lower and $1P_1^\prime$ the
higher state, and $\theta$ is given in the convention of Ref.~\cite{Godfrey:2004ya}. Reference~\cite{Martin-Gonzalez:2022qwd}
uses nearly unmixed $^3P_1$ (lower) and $^1P_1$ (higher) states and quotes no $\theta$. $^{\rm a}$The original paper
labels the lower state $1P_1^\prime$ and quotes $\theta=-24.3^\circ$. $^{\rm b}$From the mixing eigenvectors of
Ref.~\cite{S:2024wld}. $^{\rm c}$The original paper labels the lower state $1P_1^\prime$.}
\label{tab:comparison}
\begin{ruledtabular}
{\small\setlength{\tabcolsep}{4pt}
\begin{tabular}{lccccc}
Approach / Ref. & $\theta$ & $1P_1\to\Bc\gamma$ & $1P_1\to\Bcs\gamma$ & $1P_1^\prime\to\Bc\gamma$ & $1P_1^\prime\to\Bcs\gamma$ \\
\hline
Potential model~\cite{Gershtein:1994jw} & $17.1^\circ$ & 11.6 & 77.8 & 131.1 & 8.1 \\
Potential model~\cite{Fulcher:1998ka} & $28.5^\circ$ & 32.5 & 75.8 & 128 & 26.2 \\
Relativistic quark model~\cite{Ebert:2002pp} & $20.4^\circ$ & 18.4 & 78.9 & 132 & 13.6 \\
Relativized quark model~\cite{Godfrey:2004ya} & $22.4^\circ$ & 13 & 60 & 80 & 11 \\
Potential model~\cite{Eichten:2019gig} & $18.7^\circ$ & 9.9 & 62.5 & 92.3 & 7.5 \\
Nonrelativistic quark model~\cite{Li:2019tbn} & $35.5^\circ$ & 35 & 70 & 74 & 40 \\
Potential model~\cite{Li:2022bre}$^{\rm a}$ & $65.7^\circ$ & 76.5 & 13.75 & 16.5 & 71.52 \\
Constituent quark model~\cite{Martin-Gonzalez:2022qwd} & -- & $1.5\times10^{-4}$ & 146 & 173 & $1.8\times10^{-3}$ \\
Potential model~\cite{Li:2023wgq} & $35.2^\circ$ & 30.1 & 47.8 & 64.0 & 25.6 \\
Potential model~\cite{S:2024wld}$^{\rm b}$ & $43.2^\circ$ & 61 & 58 & 77 & 57 \\
Screened potential model~\cite{Bokade:2025lmn}$^{\rm c}$ & $11.8^\circ$ & 5.4 & 86.5 & 128.8 & 3.9 \\
This work (QCD SR) & $\thBc^\circ$ & $\GOneP$ & $\GOneV$ & $\GTwoP$ & $\GTwoV$ \\
\end{tabular}}
\end{ruledtabular}
\end{table*}

\section{Conclusion}
\label{sec:conclusion}

We have calculated the radiative decays of the axial-vector $B_c(1P)$ mesons into $\Bc\gamma$ and $\Bcs\gamma$
within three-point QCD sum rules, with the physical states described by the mixed $^3P_1$ and $^1P_1$ currents at the
mixing angle $\theta=\thBc^\circ$~\cite{Aliev:2026Bcmix}. In the operator product expansion, we take into account
the perturbative contribution, exact in both quark masses, and the nonperturbative contributions up to dimension six,
which change the couplings by at most about $1\%$.

We obtain $\Gamma[\Pone\to\Bc\gamma]=\GOneP~\keV$, $\Gamma[\Pone\to\Bcs\gamma]=\GOneV~\keV$,
$\Gamma[\Ptwo\to\Bc\gamma]=\GTwoP~\keV$, and $\Gamma[\Ptwo\to\Bcs\gamma]=\GTwoV~\keV$. As in the heavy-strange
sector, the contributions of the two basis currents to $g$ and $g_1$ interfere constructively for the lower state and
destructively for the higher state.

At this large mixing angle both axial-vector states decay into $\Bc\gamma$ and $\Bcs\gamma$ with comparable rates,
whereas at the mixing angles of most quark models the lower state decays mainly into $\Bcs\gamma$ and the higher one
into $\Bc\gamma$. The ratios of the two radiative widths of each state, $R_1=\ROne$ and $R_2=\RTwo$, do not depend
on the residues of the axial-vector states, and the ratio of the higher state changes by orders of magnitude over the
range of mixing angles considered in the literature. These ratios enter theory-constrained analyses of the
$\Bc\gamma$ spectrum, and a measurement of the relative yields of the $\Bc\gamma$ and $\Bcs\gamma$ lines of the two
states would test the large mixing angle directly. The same method can be applied to the radiative decays of the
$^3P_0$ and $^3P_2$ states, which complete the $B_c(1P)$ multiplet.

\appendix

\section{Perturbative spectral densities}
\label{app:pert}

Before the continuum subtraction, the double-Borel-transformed perturbative amplitudes have the following form
\begin{equation}
\widehat\Pi^{\rm pert}_X(M^2,\infty)=\sum_{j=b,c}e_j\int_0^1dx\,\Big[M^2a_{X,j}(x)+M^4b_{X,j}(x)\Big]e^{-S(x)/M^2},
\label{eq:feyn}
\end{equation}
with $S(x)$ of Eq.~\eqref{eq:G2}. The continuum is subtracted in the triangular duality region by writing
$M^2e^{-S/M^2}$ and $M^4e^{-S/M^2}$ as integrals over $s\ge S$ with the weights $e^{-s/M^2}$ and $(s-S)e^{-s/M^2}$
and restricting them to $s\le s_{\rm eff}$. Exchanging the order of the integrations gives Eq.~\eqref{eq:pert} with
\begin{equation}
\rho_{X,j}(s)=\int_{x_-(s)}^{x_+(s)}dx\,\Big[a_{X,j}(x)+b_{X,j}(x)\big(s-S(x)\big)\Big], \qquad
x_\pm(s)=\frac{s+m_b^2-m_c^2\pm\sqrt\lambda}{2s},
\label{eq:rhoint}
\end{equation}
which yields the spectral densities of Eqs.~\eqref{eq:rhoP} and \eqref{eq:FG} and, for the structures $T_1$ and $T_2$
of the transitions $A_i\to\Bcs\gamma$,
\begin{equation}
\begin{aligned}
\rho^{T_1}_A(s)&=e_b\,H(s;m_b,m_c)+e_c\,H(s;m_c,m_b), &
\rho^{T_1}_B(s)&=e_c\,K(s;m_c,m_b)-e_b\,K(s;m_b,m_c),\\
\rho^{T_2}_A(s)&=0, &
\rho^{T_2}_B(s)&=e_c\,N(s;m_c,m_b)-e_b\,N(s;m_b,m_c),
\end{aligned}
\label{eq:rhoV}
\end{equation}
with
\begin{equation}
\begin{aligned}
H(s;m_1,m_2)&=\frac{3}{4\pi^2}\Big[(1-2u_0)\sqrt\lambda+m_1(m_2-m_1+2u_0m_1)L(s;m_1,m_2)\Big],\\
K(s;m_1,m_2)&=\frac{3}{4\pi^2}\,\frac{m_1-m_2}{m_1+m_2}\Big[\big(m_2^2-m_1^2+(1-4u_0)s\big)\frac{\sqrt\lambda}{2s}
+m_1(m_2+2u_0m_1)L(s;m_1,m_2)\Big],\\
N(s;m_1,m_2)&=\frac{3}{4\pi^2}\,\frac{\bar u_0}{m_1+m_2}\Big[\big((m_1-m_2)^2(m_1+m_2)-(3m_1-m_2)s\big)\frac{\sqrt\lambda}{s^2}
+2m_1L(s;m_1,m_2)\Big],
\end{aligned}
\label{eq:HKN}
\end{equation}
and $L(s;m_1,m_2)$ of Eq.~\eqref{eq:L}.

\section{Two-point functions}
\label{app:2pt}

The Borel-transformed two-point amplitudes $\widehat\Pi_{XY}$ that enter Eq.~\eqref{eq:fA} are
\begin{equation}
\begin{aligned}
\widehat\Pi_{XY}(M^2,s_0)={}&\int_{(m_b+m_c)^2}^{s_0}ds\,e^{-s/M^2}\rho_{XY}(s)\\
&+\int_{x_-(s_0)}^{x_+(s_0)}dx\,e^{-S(x)/M^2}\sum_{n}\frac{\GGc C^{XY}_n(x)+\langle g_s^3fG^3\rangle D^{XY}_n(x)
+\langle g_s^4j^2\rangle E^{XY}_n(x)}{(M^2)^n}.
\end{aligned}
\label{eq:2ptsr}
\end{equation}
The spectral densities are
\begin{equation}
\begin{aligned}
\rho_{AA}(s)&=\frac{\sqrt\lambda}{8\pi^2}\Big[2-\frac{m_b^2+m_c^2+6m_bm_c}{s}-\frac{(m_b^2-m_c^2)^2}{s^2}\Big],\\
\rho_{AB}(s)&=\frac{3\sqrt\lambda}{8\pi^2}\,\frac{(m_b-m_c)\big[(m_b+m_c)^2-s\big]}{(m_b+m_c)s},\\
\rho_{BB}(s)&=-\frac{\sqrt\lambda}{8\pi^2}\,\frac{\big[(m_b+m_c)^2-s\big]\big(2m_b^2-4m_bm_c+2m_c^2+s\big)}{(m_b+m_c)^2s}.
\end{aligned}
\label{eq:rho2}
\end{equation}
The densities $\rho_{AA}$, $\rho_{AB}$, and $\rho_{BB}$ agree with those of Ref.~\cite{Aliev:2026Bcmix} up to an
overall normalization. The coefficient functions $C^{XY}_n$, $D^{XY}_n$, and $E^{XY}_n$ are lengthy and are
not presented here.

\bibliographystyle{utcaps_mod}
\bibliography{all}

\end{document}